\documentstyle[twocolumn,aps]{revtex}

\begin{document}
\draft
\title{Magneto-exciton in planar type II quantum dots in a perpendicular magnetic
field}
\author{K. L. Janssens\cite{karenmail}, B. Partoens\cite{bartmail} and F. M. Peeters 
\cite{peetersmail}}
\address{Departement Natuurkunde, Universiteit Antwerpen (UIA), Universiteitsplein 1,%
\\
B-2610 Antwerpen, Belgium}
\date{\today}
\maketitle

\begin{abstract}
We study an exciton in a type II quantum dot, where the electron is confined
in the dot, but the hole is located in the barrier material. The exciton
properties are studied as a function of a perpendicular magnetic field using
a Hartree-Fock mesh calculation. Our model system consists of a planar
quantum disk. Angular momentum $(l)$ transitions are predicted with
increasing magnetic field. We also study the transition from a type I to a
type II quantum dot which is induced by changing the confinement potential
of the hole. For sufficiently large magnetic fields a re-entrant behaviour
is found from $l_{h}=0$ to $l_{h}\neq 0$ and back to $l_{h}=0$, which
results in a transition from type II to type I.
\end{abstract}

\pacs{PACS: 73.21.La, 71.35.Ji, 85.35.Be}

\section{Introduction}

Self-assembled quantum dots \cite{bimbergboek} have become the subject of
intensive research, both theoretically and experimentally, since their first
realization in the early nineties \cite{eaglesham,snyder,leonard,moison}.\
The reason for this large interest is e.g. due to their possible
applications in opto-electronic devices, such as quantum dot lasers. The
formation of this type of dots by the Stranski-Krastanow growth mode
requires two semiconductor materials with a considerable lattice mismatch of
typically 5\%. Many experimental \cite{wang,polimeni,bockelmann,bayer,wilson}
and theoretical \cite{stier,brasken,xie,ulloa,karen} works are devoted to
type I structures, e.g. $InAs/GaAs$ or $InAlAs/AlGaAs,$ where both electrons
and holes are located inside the quantum dots.

Also very interesting, though yet less studied, are the type II quantum
dots, where the quantum dot forms an antidot for one of the types of
carriers, e.g. for the holes in typically the $InP/GaInP$ system or the
electrons in\ e.g. $GaSb/GaAs$. Landau level formation in strongly optically
populated type II dots was observed by Nomura {\it et al}.\ \cite{nomura} in
the photoluminescence spectra at high magnetic fields. Other
magneto-photoluminescence experiments on vertically stacked $InP$ quantum
dots were performed by Hayne {\it et al}. \cite{hayne}. Sugisaki {\it et al}%
. \cite{sugisaki} studied the magnetic field effects in a single $InP$ dot.
The optical recombination spectrum and the carrier dynamics of the $%
GaSb/GaAs $ system have been studied experimentally by Hatami {\it et al.} 
\cite{hatami}.

Whereas the type I system has been the subject of many theoretical works,
only few theoretical studies have paid attention to the type II system.
Pryor {\it et al}. \cite{pryor}\ studied the electronic structure of $%
InP/GaInP$, using a strain-dependent ${\bf k}\cdot {\bf p}$ Hamiltonian.
Also for $InP/GaInP$ dots, Nomura {\it et al}. \cite{nomura2} performed a
theoretical calculation of the Landau levels in a high magnetic field, by
solving the Hartree equations self-consistently. Using the Hartree-Fock
approximation, the binding energy of excitons, charged excitons and
biexcitons was studied by Lelong {\it et al}. \cite{lelong} in $GaSb/GaAs$
dots at zero magnetic field. The magneto-exciton in a $GaSb/GaAs$ dot was
investigated by Kalameitsev {\it et al}.\ \cite{kalameitsev}. They found
transitions of the angular momentum with increasing magnetic field.

In the present paper, we focus our attention to the properties of a single
exciton, which is bound by the Coulomb interaction in a {\it model} type II
quantum dot. We will take material parameters of the $InP/GaInP$ system.
Furthermore, we apply an external magnetic field in the growth direction,
i.e. ${\bf B}=B{\bf e}_{z}$. Including a magnetic field allows us to
investigate the transition region from exciton confinement due to the
Coulomb potential, to a confinement which is due to the magnetic field. As a
model system we take a planar quantum disk, and assume that the particles
are confined in a plane in the $z$-direction. Strain effects are neglected
in this model system.

In our model type II quantum dot the electron is confined in the dot and the
hole sits outside. The corresponding geometry is shown in Fig.~1. The
reverse confinement situation will lead to the same physics. As we do not
take the confinement effects due to strain into account, the hole is only
confined because of the Coulomb attraction to the electron. As we have no a
priori knowledge about the width of the hole wavefunction it is difficult to
choose good basisfunctions for the expansion of the hole wavefunction.
Therefore, we solved the Hartree-Fock (HF) equations on a grid, which allows
very flexible solutions, in principle of arbitrary shape. With the same
motivation, similar Hartree-Fock mesh calculations were recently used in
atomic physics~\cite{ivanov}.

As confinement potential, we take hard walls of finite height. By varying
the hole confinement potential, we can study the transition from a type I
structure (i.e., the hole is confined in the dot) to a type II structure
(i.e., the dot is a barrier for the hole). We show that for small antidots
the attraction of the hole to the electron is stronger than the barrier
energy and the system is still type I. Increasing the barrier height and/or
the size of the dot induces a transition to a type II system. Furthermore,
we found angular momentum transitions with increasing magnetic field. For
large enough magnetic fields (depending on the height of the potential
barrier), we find a new re-entrant behaviour to the zero angular momentum
state.

The paper is organized as follows. In Sec. II, we describe briefly our
theoretical model. The numerical results are presented in Sec. III. In Part
A of this Section, we discuss the effect of a varying magnetic field and
explain the origin of the angular momentum transitions. Part B deals with
the transition from a type I to a type II system. Part C is dedicated to the
re-entrant behaviour. In the last Part, D, we present the results for the
excitation spectrum. Our results are summarized in Sec. IV. In the Appendix,
we discuss in more detail the method we used for the calculation of the
Hartree integral.

\section{Theoretical model}

The energies and wavefunctions are obtained by solving the following HF
single particle equations in the effective mass approximation (with $m_{e}$
and $m_{h}$ the effective electron and hole masses, respectively, $r_{e,h}=%
\sqrt{x_{e,h}^{2}+y_{e,h}^{2}},$ $\omega _{c,e}=eB/m_{e}$ and $\omega
_{c,h}=eB/m_{h}$) 
\begin{eqnarray}
\left[ -\frac{\hbar ^{2}}{2m_{e}}\frac{1}{r_{e}}\frac{\partial }{\partial
r_{e}}\left( r_{e}\frac{\partial }{\partial r_{e}}\right) +\frac{\hbar ^{2}}{%
2m_{e}}\frac{l_{e}^{2}}{r_{e}^{2}}+\frac{l_{e}}{2}\hbar \omega _{c,e}+\frac{1%
}{8}m_{e}\omega _{c,e}^{2}r_{e}^{2}\right. &&  \nonumber \\
\left. +V_{e}(r_{e})-\frac{e^{2}}{4\pi \epsilon }\int \frac{\rho
_{h}(r^{\prime })}{|{\bf r}-{\bf r}^{\prime }|}d{\bf r}^{\prime }\right]
\psi _{e}(r_{e})=\epsilon _{e}\psi _{e}(r_{e}), &&  \eqnum{1a} \\
\left[ -\frac{\hbar ^{2}}{2m_{h}}\frac{1}{r_{h}}\frac{\partial }{\partial
r_{h}}\left( r_{h}\frac{\partial }{\partial r_{h}}\right) +\frac{\hbar ^{2}}{%
2m_{h}}\frac{l_{h}^{2}}{r_{h}^{2}}-\frac{l_{h}}{2}\hbar \omega _{c,h}+\frac{1%
}{8}m_{h}\omega _{c,h}^{2}r_{h}^{2}\right. &&  \nonumber \\
\left. +V_{h}(r_{h})-\frac{e^{2}}{4\pi \epsilon }\int \frac{\rho
_{e}(r^{\prime })}{|{\bf r}-{\bf r}^{\prime }|}d{\bf r}^{\prime }\right]
\psi _{h}(r_{h})=\epsilon _{h}\psi _{h}(r_{h}), &&  \eqnum{1b}
\end{eqnarray}
\noindent where we made use of the axial symmetry by taking $\Psi
_{e}(r_{e},\varphi _{e})=e^{il_{e}\varphi _{e}}\psi _{e}(r_{e})$ and $\Psi
_{h}(r_{h},\varphi _{h})=e^{il_{h}\varphi _{h}}\psi _{h}(r_{h}),$ and where
the densities $\rho _{e}(r^{\prime })$ and $\rho _{h}(r^{\prime })$ are
given by respectively $\left| \Psi _{e}(r_{e},\varphi _{e})\right| ^{2}$ and 
$\left| \Psi _{h}(r_{h},\varphi _{h})\right| ^{2}$. The Hartree-fock
equations were solved using a finite difference scheme. More details about
the implementation of this finite difference scheme can be found in Refs.~ 
\cite{peeters,karen}. Note that there are no exchange terms as we only
consider a single electron and a single hole. However, these equations can
still be called HF as the self-interaction is excluded. As confinement
potentials we take hard walls of finite height: 
\begin{equation}
V_{e,h}(r_{e},r_{h})=\left\{ 
\begin{array}{c}
V_{e,h},\;\;r_{e,h}>R, \\ 
0,\;\;\mbox{otherwise},
\end{array}
\right.  \eqnum{2}
\end{equation}
with $R$ the radius of the disk, and where we took $V_{e}$ positive and $%
V_{h}$ negative. Note that the only good quantum number is the total angular
momentum in the $z$-direction, defined by $L=l_{e}+l_{h}.$

These equations must be solved self-consistently, which is done iteratively.
We start with the free electron solution because in the absence of any
Coulomb interaction only the free electron is confined. The Hartree
integrals are integrated numerically 
\begin{equation}
\int \frac{\rho (r^{\prime })}{|{\bf r}-{\bf r}^{\prime }|}d{\bf r}^{\prime
}=4\int \frac{\rho (r^{\prime })r^{\prime }}{r+r^{\prime }}{\cal K}\left( 
\frac{4rr^{\prime }}{(r+r^{\prime })^{2}}\right) dr^{\prime },  \eqnum{3}
\end{equation}
where ${\cal K}(x)$ is the complete elliptic integral of the first kind.
More details about the calculation and numerical implementation of this
integral is given in the Appendix.

After convergence, the total energy is given by 
\begin{equation}
E_{\mbox{exciton}}=\epsilon _{e}+\epsilon _{h}+\frac{e^{2}}{4\pi \epsilon }%
\int \int \frac{\rho _{e}(r)\rho _{h}(r^{\prime })}{|{\bf r}-{\bf r}^{\prime
}|}d{\bf r}d{\bf r}^{\prime }.  \eqnum{4}
\end{equation}
The contribution of the correlation to the total energy is neglected in HF,
but for the self-assembled quantum dots, it is expected to be less than 2\% 
\cite{brasken}.

\section{Results}

\subsection{Angular momentum transitions}

First, we calculated the groundstate energy of the exciton as a function of
the external magnetic field. We took the following parameters: $%
m_{e}=0.077m_{0},$ $m_{h}=0.6m_{0},$ $V_{e}=250meV,$ $V_{h}=-50meV,$ and $%
\epsilon =12.61,$ which are typical for the $InP/GaInP$ system \cite{pryor}
and consider a dot of radius $R=8nm$ \cite{hayne}. Our numerical results are
depicted in Fig.~2 and show that the exciton groundstate exhibits
transitions of the angular momentum $l_{h}$ as a function of the magnetic
field (indicated by the arrows). These changes in angular momentum of the
groundstate are {\it not} present in type I\ dots and are a direct
consequence of the fact that we are dealing with type II dots. For the hole,
the disk acts as a barrier and by increasing the magnetic field the hole is
pushed closer to the disk boundary which leads to an increase of the hole
potential energy. For a certain magnetic field it is for the hole
energetically more favourable to jump to a higher $l_{h}$ state, which
brings the hole further away from the disk interface. This is also
demonstrated in the inset of Fig.~2, where a contourplot of the density of
the hole wavefunction is shown, as a function of both the magnetic field and
the radial position. It is apparent that the hole is located close to the
disk, even at zero magnetic field and that with increasing magnetic field
the hole is pushed closer to the border of the disk. At the angular momentum
transitions the hole is spread out a little more, and jumps a distance away
from the disk interface. But note that on the average the hole is pushed
closer to the disk boundary and its width decreases with increasing magnetic
field. For the present case we find five transitions for a magnetic field up
to $B=50T$.

The magnetic field values at which the angular momentum transitions occur
will depend on the disk radius $R$. Fig.~3 shows a phase diagram of the $%
l_{h}$ transitions as a function of the magnetic field $B$ and the disk
radius $R,$ for the $InP$ parameters used above. With increasing disk
radius, the transitions shift to lower magnetic field values. This can be
understood as follows: for larger disks, a smaller magnetic field is needed
to push the hole close to the border of the disk, and to induce an angular
momentum transition.

Another way to understand the angular momentum transitions is as follows.
The hole is spatially confined into a ring-like area, and if we make the
extreme simplification of a zero width ring, the hole energy is given by 
\begin{equation}
E_{h}=\frac{\hbar ^{2}}{2mR^{2}}\left( l_{h}-\frac{\phi }{\phi _{0}}\right)
^{2},  \eqnum{5}
\end{equation}
from which it is clear that the ground state exhibits angular momentum
transitions each time the flux through the ring $\phi $ equals $%
(l_{h}+1/2)\phi _{0}$ with $\phi _{0}=hc/e$ the quantum of flux.

\subsection{Type II to type I transitions}

In Fig.~4 we give a closer look at the hole wavefunction for a very small
disk $R=2nm,$ $V_{h}=-50meV$ and $B=0T,$ and we find that, even without a
magnetic field, the hole is partially situated inside the quantum disk. This
is a remarkable effect, as from the shape of the effective potential,
defined by the sum of the confinement potential and the Hartree potential
(inset of Fig.~4), we would expect the hole wavefunction to be situated in
the barrier. We attribute this effect to a kind of tunneling of the hole
through the quantum disk, as a consequence of the very small disk radius.
Remark that this state continues to be the groundstate and that the effect
becomes even stronger for higher fields. Another consequence of this effect
is a higher overlap of the electron and hole wavefunctions, which is an
indication of type I behaviour.

In a next step, we studied the exciton properties as a function of the hole
confinement potential, which allows us to explore the transition region from
type I systems $\left( V_{h}>0\right) $ to type II systems $\left(
V_{h}<0\right) .$ Hereby we kept the disk radius fixed at $R=8nm.$ Fig.~5
shows the phase diagram for the angular momentum transitions as a function
of the confinement potential $V_{h}$ and the magnetic field $B.$ A feature
that immediately catches the eye is that up to $V_{h}\simeq -24.5meV$ the $%
l_{h}=0$ state remains the groundstate over the total $B$-region under
consideration. Investigating this more in depth, we find that, even for $%
B=0T,$ the hole wavefunction is located almost entirely inside the quantum
disk. This is a consequence of the Hartree potential (due to the attraction
to the electron) which overcomes the potential barrier of the disk.
Therefore, we can speak of type I systems up to $V_{h}\simeq -24.5meV.$

In order to have a more physical idea of the origin the type I to type II
transition, we developed the following intuitive picture. When the system is
type I, the hole will be located inside the quantum disk. This will happen
when the effective potential (the sum of the hole confinement potential and
the Hartree potential, i.e. $V_{c}(r_{h})+V_{h}$ (see the inset of Fig.~4))
is lower at $r_{h}=0$ than at $r_{h}=R.$ For a type II system however, the
effective potential will be lower at the boundary, and the hole will prefer
to sit outside the disk but near the boundary.\ Generally, we can state that
the transition from type I to type II occurs when the effective potential at
the origin equals the one at the radial boundary. Because we take the
confinement potential zero inside the disk, this leads to the following
formula: 
\begin{equation}
V_{c}(r_{h}=0)=V_{h}+V_{c}(r_{h}=R).  \eqnum{6}
\end{equation}
From this equality, we can find an estimate for the confinement potential $%
V_{h}$ at which the type I to type II transition occurs. The Hartree
potential was calculated within the approximation of an infinitely high
electron confinement potential, where the electron wavefunction can be
expressed by the Bessel functions $J_{0}$, i.e. 
\begin{equation}
V_{c}(r_{h})=-\frac{e^{2}}{4\pi \epsilon }N^{2}\int \frac{\left|
J_{0}(kr_{e}^{\prime })\right| ^{2}}{\left| r_{h}-r_{e}^{\prime }\right| }d%
{\bf r}_{e}^{\prime },  \eqnum{7}
\end{equation}
with $N^{2}=2/(R^{2}\left| J_{1}(kr_{e}^{\prime })\right| ^{2})$ the
normalization, and $k=\sqrt{2m_{e}E}/\hbar $. In Fig.~6, we plot the values
for $V_{h}$ as a function of the disk radius $R$ at which such a type I to
type II transition occurs. We show the results obtained by the full
Hartree-Fock calculation (full curve, squared dots)\ and the approximated
results, as obtained from Eq.~(6) (dashed curve, circular dots). We find
that for large $R$ the two curves converge to each other, but that for small 
$R$ an appreciable discrepancy exists. This is due to the fact that for
small disk radii, the approximation of a hard wall confinement is less
justified. By pushing the electron wavefunction completely into the disk (in
contrast to the `real' case of a finite potential, where the wavefunction
can tunnel into the barrier), the Hartree potential is strongly enhanced,
thereby leading to a strong enhancement of the critical confinement
potential $V_{h}.$

Another interesting property is the probability for recombination of the
exciton. This is proportional to the square of the overlap integral \cite
{kalameitsev} 
\begin{eqnarray}
I &=&\int \Psi _{e}\left( {\bf r}\right) \Psi _{h}\left( {\bf r}\right) d%
{\bf r}  \nonumber \\
&=&\int_{0}^{2\pi }e^{i(l_{e}+l_{h})}d\varphi \int_{0}^{\infty }\psi
_{e}\left( r\right) \psi _{h}\left( r\right) rdr\text{ .}  \eqnum{8}
\label{eq6}
\end{eqnarray}
Notice that the first integral is equal to $2\pi \delta _{l_{e}+l_{h}},$
what means that the probability for de-excitation is only non-zero for the
case $l_{e}+l_{h}=0.$ This implies that after an angular momentum transition
the probability for recombination of an exciton decreases drastically. In
photoluminescence (PL) experiments, one will observe a strong quenching or
even disappearance of the PL spectrum after a certain value of the magnetic
field.

Fig.~7 shows the overlap integral $I$ as a function of $V_{h},$ for $B=0T.$
Without a magnetic field, the $l_{h}=0$ state is always the groundstate and
therefore, $I$ is non-zero over the total region. Up to $V_{h}=-25meV$ the
overlap is large, and further increasing $-V_{h},$ we find a sudden strong
decrease of the overlap. The reason for this behaviour is directly related
to the position of the hole wavefunction. As long as the hole is sitting
inside the disk, the overlap will be very large. However, from the moment
the hole jumps outside the disk, the overlap decreases strongly. In fact,
from Fig.~7 we can infer immediately the position of the hole. Furthermore,
the region of the strong decrease in overlap indicates the transition from
type I to type II behaviour. The dashed line gives the overlap integral at $%
B=20T.$ As we see from Fig.~5, a transition to the $l_{h}=1$ state occurs
when the confinement potential $V_{h}$ approaches $-24.5meV$ and the
condition $l_{e}+l_{h}=0$ for recombination of the exciton, is no longer
satisfied, which leads to $I=0.$ The recombination of the exciton will
happen through indirect processes, resulting in a much longer lifetime of
the exciton. The changing lifetime can be detected experimentally by a
changing lineshape \cite{yasuhira,schubert}.

\subsection{Re-entrant behaviour}

In this section, we concentrate more closely on the type I - type II
transition region, i.e. $V_{h}$ between $-20meV$ and $-30meV.$ As an
example, we investigated the exciton groundstate energy for $V_{h}=-27meV.$
The result is depicted in Fig.~8 which shows one additional remarkable
feature: after several $l_{h}$ transitions with increasing magnetic field,
we find at sufficiently large magnetic field, i.e. $B\simeq 42T,$ a
re-entrance of the $l_{h}=0$ state. It is interesting also to take a look at
the evolution of the wavefunction with increasing magnetic field. This is
depicted as a contourplot in the inset of Fig.~8. Initially, for very small
magnetic fields, we find that a small part of the wavefunction has already
entered the dot region. However, at $B\simeq 6T,$ due to a jump to a higher
angular momentum state, the hole wavefunction is pushed outside the dot
region. Further increasing the magnetic field leads to more $l_{h}$
transitions, as we found already in Section III.A. At the specific magnetic
field value however, where the re-entrance of $l_{h}=0$ occurs, we find that
suddenly the hole wavefunction jumps almost entirely inside the disk. At
this point, the magnetic field and the attraction of the electron overcomes
the potential barrier of the quantum disk, and it will be energetically more
favourable for the hole to sit inside the disk.

The re-entrant behaviour is also visible in the $\left( B,V_{h}\right) $%
-phase diagram (Fig.~5). We want to emphasize that there will be a
re-entrant behaviour for any value of $V_{h}<-24.5meV$, for sufficiently
large magnetic fields. This can already be seen from Fig.~5, where the line
which indicates the transition from a certain $l_{h}$-state to the $l_{h}=0$
state is not a straight vertical line, but has a small slope. For example
for $V_{h}=-50meV$ we found that a magnetic field of $B=193T$ is needed to
induce this re-entrance to the $l_{h}=0$ state.

Another question which arises is how the disk radius influences the
re-entrant behaviour. In our previous investigation of the influence of the
disk radius on the groundstate energy, we found no evidence of this, because
we did not consider large enough magnetic fields for the confinement
potential under consideration $\left( V_{h}=-50meV\right) .$ Therefore, we
decided to make a new $\left( B,R\right) $-phase diagram, this time for $%
V_{h}=-25meV,$ for which we know from Fig.~5 that re-entrant behaviour
occurs at rather small magnetic fields. The result is depicted in Fig.~9,
and the first striking feature is that the re-entrant behaviour occurs for
any disk with radius $R>8nm.$ Furthermore, we find more $l_{h}$ transitions
for larger disk radii, hereby increasing the magnetic field at which the
re-entrance of $l_{h}=0$ takes place. However at $R\simeq 12nm,$ we find
that the magnetic field position of the re-entrant behaviour reaches a
maximum value, $B\simeq 42T.$ For larger disk radii, the re-entrance occurs
at slightly decreasing magnetic field. Indeed, with increasing disk radius,
electron and hole are drawn more and more apart, and therefore it will
sooner become energetically more favourable for the hole to jump inside the
disk.

This re-entrant behaviour can be understood qualitatively from the following
simple model. We compare the approximate energies for a hole of respectively
a type II and a type I system: 
\begin{eqnarray}
E_{h}^{II} &=&\frac{3}{2}\hbar \omega _{c,h}-\frac{e^{2}}{4\pi \epsilon
_{0}\epsilon }\frac{1}{R}-V_{h},  \eqnum{9a} \\
E_{h}^{I} &=&\frac{1}{2}\hbar \omega _{c,h}-E_{Coulomb}.  \eqnum{9b}
\end{eqnarray}
The approximate energy for a hole in a type II system (Eq.~(9a)) is
constructed by approximating the disk (with radius $R$) by an infinitely
high barrier for the hole. The first term of Eq.~(9a) gives the one-particle
energy of the hole, which is just the first energy level of a particle in a
magnetic field which fulfils the zero-wavefunction condition at the disk
boundary. The second term is the Coulomb interaction between the electron
and the hole, and the third is the potential energy of the hole. In a type I
system (Eq.~(9b)) both electron and hole are located inside the disk, and
are subjected to a magnetic field. Now the one-particle hole energy is just
the energy of the first Landau level. We approximated the Coulomb energy
(second term in Eq.~(9b)) by using the single particle wavefunction of
electron and hole. At strong magnetic fields and for large disk radii i.e. $%
R>l_{B}$, the potential confinement by the disk can be neglected with regard
to the confinement by the magnetic field. The single particle wavefunctions
are then the well-known wavefunctions of a particle in a magnetic field,
given by 
\begin{equation}
\varphi _{e,h}\left( {\bf r}_{e,h}\right) =\sqrt{\frac{1}{\pi }}\frac{1}{%
\sqrt{2}l_{B}}\exp \left( -\frac{r_{e,h}^{2}}{4l_{B}^{2}}\right) , 
\eqnum{10}
\end{equation}
for the groundstate $(n=0,l_{h}=0).$ As the wavefunctions are
mass-independent, there is no distinction between the electron and the hole.
Therefore we can treat our system as being completely analogous to a system
consisting of two electrons in a magnetic field, for which our first order
approximation of the Coulomb interaction energy can be calculated
analytically, and which reduces to 
\begin{eqnarray}
E_{Coul} &=&\frac{e^{2}}{4\pi \epsilon _{0}\epsilon }\frac{\sqrt{\pi }}{2}%
\sqrt{\frac{e}{\hbar }}\sqrt{B}  \eqnum{11a} \\
&=&C_{coul}\sqrt{B}.  \eqnum{11b}
\end{eqnarray}

Equalization of Eqs.~(9a) and (9b) gives us the magnetic field at which the
re-entrance occurs as a function of both disk radius $R$ and confinement
potential $V_{h}.$ The transition magnetic field can be obtained
analytically as 
\begin{eqnarray}
B &=&\frac{m_{h}^{2}}{2e^{2}\hbar ^{2}}C_{coul}\times \left( C_{coul}-\sqrt{%
C_{coul}^{2}+\frac{4e\hbar }{m_{h}}\left( \frac{e^{2}}{4\pi \epsilon
_{0}\epsilon }\frac{1}{R}+V_{h}\right) }\right)   \nonumber \\
&&+\frac{m}{e\hbar }\left( \frac{e^{2}}{4\pi \epsilon _{0}\epsilon }\frac{1}{%
R}+V_{h}\right) .  \eqnum{12}
\end{eqnarray}

For a fixed radius $R,$ we can vary $V_{h}$ and deduce the magnetic field at
which the transition occurs. The result is shown by the dashed curve in
Fig.~5. For large fields, the approximated curve (dashed line) has
qualitatively the same behaviour as the curve obtained from the full
Hartree-Fock treatment and the two curves converge to each other. The result
for a fixed $V_{h},$ when varying the disk radius $R,$ is shown by the
dashed curve in Fig~9. We find a perfect agreement for large disk radii,
where our model is valid. The discrepancy for small disk radii is a
consequence of our assumption to neglect the disk confinement.

Fig.~10 shows the overlap integral $I$ as a function of the magnetic field,
for confinement potentials of the hole $V_{h}=-25meV$ (solid curve) and $%
V_{h}=-27meV$ (dashed curve) and a fixed disk radius $R=8nm.$ At first we
find a slowly increasing overlap, which is a consequence of the increasing
magnetic field, pushing the particles closer together. The already rather
large value of the overlap indicates that a considerable part of the hole is
already situated inside the disk. When the first $l_{h}$-transition occurs,
the overlap falls immediately down to zero, because $l_{e}+l_{h}\neq 0$. The
overlap remains zero, until the $l_{h}=0$ state returns as the groundstate
and the condition $l_{e}+l_{h}=0$ is satisfied. This re-entrance of $l_{h}=0$
is accompanied by a jump of the wavefunction in the disk, and this leads to
the strong enhancement of the overlap value. We see that the re-entrance of
the $l_{h}=0$ state happens at lower magnetic fields for the lower potential
barrier.

Fig.~11 shows the overlap integral $I$ as a function of the disk radius $R$
for $B=0T$ (solid curve), $B=20T$ (dashed curve) and $B=40T$ (dotted curve).
This figure gives evidence for the fact that for very small radii the hole
wavefunction is almost entirely situated inside the disk. For increasing
disk radius, the hole is pushed more and more outside the disk, hereby
decreasing the value of the overlap integral. This decrease is initially
less for increasing magnetic field because of the enhanced localization
effect. For sufficiently large magnetic fields, $l_{h}$ transitions are
induced, which leads to a zero overlap integral. Also here we see that for
sufficient large $R$ a re-entrant behaviour to the $l_{h}=0$ state is found
at which point the overlap integral becomes again non-zero.

\subsection{Excitation spectrum}

Lastly, we investigated the exciton energy spectrum as a function of the
magnetic field. The physical parameters used in this calculation are the
ones mentioned in Section II, with a disk radius $R=8nm.$ We considered
states with different radial quantum numbers $k_{e},$ $k_{h}$ and different
angular momenta $l_{e}$, $l_{h}.$ Note that $k_{e}$ and $k_{h}$ are
approximate quantum numbers for the current system. Furthermore, our results
are a {\it first order }perturbation theory approximation to the real energy
spectrum, as we perform only one HF iteration, i.e. the energy of the
exciton is obtained by solving the equation for the hole in the field of the
confined electron. Note that changing the electron quantum number results in
a strong increase of the energy value. The states $\left( k_{e},l_{e}\right)
=\left( 2,0\right) $ and $\left( 1,\pm 2\right) $ appear to be already
unbound, i.e. the energy exceeds the electron barrier of $250meV$. The inset
of Fig.~12(a) shows the bound states of the energy spectrum where we varied
both $k_{e}$ and $l_{e},$ keeping $k_{h}$ and $l_{h}$ fixed at $\left(
0,0\right) .$ The main part of Fig.~12(a) shows the energy spectrum for
fixed $\left( k_{e},l_{e}\right) =\left( 0,0\right) $ and varying the hole
quantum numbers $k_{h}$ and $l_{h}.$ We find that now the energy values span
a smaller energy region. This is due to the fact that: i) the hole is much
heavier than the electron, and therefore has substantially lower energies,
and ii) the hole is less confined. In fact, for every possible value of the
electron quantum numbers $k_{e}$ and $l_{e},$ one has a spectrum of all
possible $\left( k_{h},l_{h}\right) $ values, and because these span a
smaller energy region, the total energy spectrum will consist of mainly the
electron branches, with superimposed on each of them the spectrum with the
changing hole quantum numbers.

Notice also the anomalous behaviour of certain states, e.g. $\left(
k_{h},l_{h}\right) =\left( 2,0\right) $ in the high magnetic field region.
To investigate this further, we concentrated on the variation of the radial
quantum number $k_{h},$ keeping the angular momentum $l_{h}$ fixed at 0.
This result is shown in Fig.~12(b) which clearly shows the occurrence of
anti-crossings. These anti-crossings are due to the fact that the radial
quantum number is not a good quantum number, leading to strong mixing of
radial states at the anti-crossings.

\section{Conclusions}

We investigated the exciton properties in a strongly simplified type II
model quantum disk, with the hole located in the barrier. Strain effects
were disregarded and a flat disk geometry was assumed. Because in our model
system there is no geometrical confinement for the hole, the only
``confinement'' comes from the attraction to the electron, i.e. the Coulomb
interaction energy. We solved this problem by using a Hartree-Fock mesh
calculation, which allowed us to calculate the exciton energy, without an a
priori knowledge of the single particle hole wavefunction.

We studied the influence of a perpendicular applied magnetic field, and
found {\it angular momentum transitions} with increasing magnetic field.
These are a consequence of the fact that the magnetic field pushes the hole
closer to the disk, making it energetically more favourable to jump to a
higher $l_{h}$ state. Varying the disk radius showed that the transitions
shift to lower magnetic field for larger $R.$ We also found that the hole is
located almost entirely inside the disk for very small disk radii.

Furthermore we investigated the transition region from {\it type I to type II%
} systems, by varying the confinement potential of the hole, $V_{h}.$ A
striking feature here is the fact that we are dealing with type I systems up
to $V_{h}>-24.5meV.$ Even at $B=0T,$ the Coulomb attraction overcomes the
potential barrier and the hole is situated inside the quantum disk. Taking a
closer look at the transition region between type I and type II, showed the
existence of a {\it re-entrant behaviour} of the $l_{h}=0$ state. This
re-entrant behaviour is coupled with a sudden jump of the wavefunction into
the disk.

The angular momentum transitions and the re-entrant behaviour should be
measurable experimentally by quenching of luminescence and/or changing
lineshapes.

In a last part, we studied the excitation spectrum as a function of the
magnetic field. We varied the quantum numbers $k$ and $l$ for both electron
and hole, and found that for every value of $\left( k_{e},l_{e}\right) $ one
has a spectrum consisting of the different radial and angular momentum hole
states. Furthermore, taking a closer look to the varying $k_{h}$ states,
with fixed $l_{h}$ and fixed electron quantum numbers, we found an
anti-crossing of levels, a consequence of the fact that $k_{h}$ is not a
good quantum number and therefore lifts the degeneracy.

\section{Acknowledgments}

K. L. J. is supported by the ``Instituut voor de aanmoediging van Innovatie
door Wetenschap en Technologie in Vlaanderen'' (IWT-Vl) and B. P. is a
post-doctoral researcher with the Flemish Science Foundation (FWO-Vl.).
Discussions with M. Hayne, M. T\u{a}di\'{c} and A. Matulis are gratefully
acknowledged. Part of this work was supported by the FWO-Vl, IUAP-IV, the
``Bijzonder Onderzoeksfonds van de Universiteit Antwerpen'' (GOA), and the
EC-project NANOMAT.

\section{Appendix: Calculation of the Hartree integral}

The Hartree integral, which expresses the effect of one particle, e.g. the
electron, on the other particle, e.g. the hole, is given by 
\begin{equation}
\int \frac{\rho (r^{\prime })}{|{\bf r}-{\bf r}^{\prime }|}d{\bf r}^{\prime
}=\int dr^{\prime }r^{\prime }\int d\varphi ^{\prime }\frac{\rho (r^{\prime
})}{\sqrt{r^{2}+r^{\prime 2}-2rr^{\prime }\cos (\varphi -\varphi ^{\prime })}%
}.  \eqnum{A1}
\end{equation}
As we are dealing with cylindrical symmetry, we can remove the $\varphi $%
-dependence. The integral over the angle becomes the complete elliptic
integral of the first kind, which converts (A1) into 
\begin{equation}
4\int \frac{\rho (r^{\prime })r^{\prime }}{r+r^{\prime }}{\cal K}\left( 
\frac{4rr^{\prime }}{(r+r^{\prime })^{2}}\right) dr^{\prime }.  \eqnum{A2}
\end{equation}
The radial integral has to be solved numerically. We use a polynomial
approximation for the elliptic function \cite{abramowitz}, namely 
\begin{equation}
{\cal K}(x)=[a_{0}+a_{1}x^{\prime }+a_{2}x^{\prime 2}]-[b_{0}+b_{1}x^{\prime
}+b_{2}x^{\prime 2}]\ln (x^{\prime }),  \eqnum{A3}
\end{equation}
with $x^{\prime }=1-x$ and where the coefficients $a_{i}$ and $b_{i}$ are
given in Ref.~\cite{abramowitz}. Since this implies the appearance of a
logaritmic divergence in the integrand, the commonly used trapezoidal rule
will give bad results. Therefore we used another method, the so-called
`logaritmically weighted method' which takes into account this problem.

Generally, the following integral can be considered: 
\begin{equation}
I(r)=\int_{0}^{1}dxF(x)\ln \frac{(x-r)^{2}}{(x+r)^{2}},  \eqnum{A4}
\end{equation}
which, after transformation, becomes 
\begin{equation}
I(r)=\sum_{i=0}^{N-1}\int_{0}^{h}dxF(x+hi)\ln \frac{(x-(r-hi))^{2}}{%
(x+(r+hi))^{2}},  \eqnum{A5}
\end{equation}
with $h$ the discretization step and $N$ the number of steps. If we replace $%
F(x+hi)$ by $F_{i}+[F_{i+1}-F_{i}]\left( x/h\right) ,$ we can write (A5) as 
\begin{equation}
I(r)=\sum_{i=0}^{N-1}\{F_{i}A_{i}(r)+[F_{i+1}-F_{i}]C_{i}(r)\},  \eqnum{A6}
\end{equation}
and the remaining problem is the calculation of the coefficients $A_{i}(r)$
and $C_{i}(r).$ The integrals which determine the coefficients can be solved
exactly, which leads to the following results: 
\begin{equation}
A_{i}(r)=a(h-(r-hi))-a(h+(r+hi)),  \eqnum{A7}
\end{equation}
with 
\begin{equation}
a(y)=2y\ln y+2(h-y)\ln (h-y),  \eqnum{A8}
\end{equation}
and 
\begin{equation}
C_{i}(r)=h^{-1}[c(h-(r-hi))-c(h+(r+hi))]-2r,  \eqnum{A9}
\end{equation}
with 
\begin{equation}
c(y)=\left( y(2h-y)\right) \ln y+(h-y)^{2}\ln \left( h-y\right) . 
\eqnum{A10}
\end{equation}

\bigskip

\begin{figure}[tbp]
\caption{The geometry of the system under consideration, with side and top
view.}
\label{Fig1}
\end{figure}

\begin{figure}[tbp]
\caption{The exciton energy as a function of the magnetic field, for $R=8nm,$
$V_{e}=250meV$ and $V_{h}=-50meV$. The successive $l_{h}$-transitions are
indicated by arrows. The inset shows a contourplot of the square of the hole
wavefunction as a function of the magnetic field.}
\end{figure}

\begin{figure}[tbp]
\caption{Phase diagram of the $l_{h}$ transitions as a function of magnetic
field $B$ and disk radius $R,$ for $V_{h}=-50meV.$}
\end{figure}

\begin{figure}[tbp]
\caption{Hole wavefunction for $R=2nm$ and $V_{h}=-50meV,$ at $B=0T.$ The
inset shows the effective confinement potential for the hole.}
\end{figure}

\begin{figure}[tbp]
\caption{Phase diagram of the successive $l_{h}$ states for varying
confinement potential $V_{h},$ as a function of the magnetic field. The disk
radius is fixed at $R=8nm.$ The dashed line indicates the result obtained by
the approximate model.}
\end{figure}

\begin{figure}[tbp]
\caption{Confinement potential of the hole at which a transition from a type
I to a type II system occurs, as a function of the disk radius $R.$ The
solid curve (squared dots) indicates the result obtained within the HF
treatment, whereas the dashed curve (circular dots)\ the approximated result
is. }
\end{figure}

\begin{figure}[tbp]
\caption{Overlap integral for a varying confinement potential of the hole,
for $R=8nm$ and at $B=0T$ (solid curve) and $B=20T$ (dashed curve).}
\end{figure}

\begin{figure}[tbp]
\caption{Exciton groundstate energy for $V_{h}=-27meV.$ The successive $%
l_{h} $ states are indicated by arrows. Notice the re-entrant behaviour of
the $l_{h}=0$ state.}
\end{figure}

\begin{figure}[tbp]
\caption{($B,R$) Phasediagram for $V_{h}=-25meV.$ The re-entrant behaviour
slowly decreases for large $R.$ The dashed curve indicates the result
obtained by the approximate model.}
\end{figure}

\begin{figure}[tbp]
\caption{Overlap integral as a function of the magnetic field, for fixed $%
R=8nm$ and for $V_{h}=-25meV$ (solid curve) and $V_{h}=-27meV$ (dashed
curve). When $l_{e}+l_{h}\neq 0,$ the overlap is 0.}
\end{figure}

\begin{figure}[tbp]
\caption{Overlap integral as a function of the disk radius, for $V_{h}=-25meV
$ and magnetic fields of respectively $0T$ (solid curve), $20T$ (dashed
curve) and $40T$ (dotted curve).}
\end{figure}

\begin{figure}[tbp]
\caption{(a) Energy spectrum for different $k_{h}$ and $l_{h},$ for fixed $%
(k_{e},l_{e})=(0,0),$ as a function of the magnetic field. Inset: idem, but
now for different $k_{e}$ and $l_{e},$ and for $(k_{h},l_{h})=(0,0).$ (b)
The same as in (a), but now for $l_{h}=0$. Notice the anti-crossings as a
function of the magnetic field, which is due to the lifting of the
degeneracy, or a strong mixing of the radial states.}
\end{figure}

\end{document}